\documentclass[useAMS,usenatbib,usegraphicx]{mn2e}
\usepackage{graphicx}
\usepackage{color}
\newcommand{\be}{\begin{equation}}
\newcommand{\ee}{\end{equation}}

\newcommand{\bl}[1]{\mbox{\boldmath$ #1 $}}

\newcommand{\degree}{\mbox{$^{\circ}$}}

\begin{document}

%\title[SEDs of fragmenting disks]{Spectral energy distributions of fragmenting protostellar disks}
\title[Properties of fragmenting protostellar discs]{Fragmenting protostellar disks: properties and observational signatures}
\author[Vorobyov et al.]{Eduard I. Vorobyov$^{1,2}$\thanks{E-mail:eduard.vorobiev@univie.ac.at (EIV)},
 Olga V. Zakhozhay$^3$\thanks{E-mail:zkholga@mail.ru (OVZ)}, and Michael M. Dunham$^{4}$ \\
$^{1}$ Institute of Astrophysics, University of Vienna, Vienna, 1180, Austria \\
$^{2}$ Research Institute of Physics, Southern Federal University, Rostov-on-Don, 344090, Russia \\
$^{3}$ Main Astronomical Observatory, National Academy of Sciences of Ukraine, Kyiv, 03680, Ukraine \\
$^{4}$ Department of Astronomy, Yale University, P.O. Box 208101, New Haven, CT 06520, USA \\
}

\maketitle

\begin{abstract}
Using numerical hydrodynamic simulations, we study the gravitational fragmentation of 
an unstable protostellar disc formed during the collapse
of a pre-stellar core with a mass  of 
$1.2~M_\odot$. The forming fragments span a mass range from 
about a Jupiter  mass
to very-low-mass protostars and are located at distances from a few tens to a thousand~AU, with
a dearth of objects at $\la 100$~AU. We explore  the possibility of observational detection of the fragments
in discs viewed through the outflow cavity at a distance of 250~pc. 
We demonstrate that one hour of integration time 
with the Atacama Larger Millimeter/sub-millimeter Array (ALMA) is sufficient to detect 
the fragments with masses as low as $1.5~M_{\rm Jup}$ at orbital distances up to 
800~AU from the protostar. The ALMA resolution sets 
the limit on the minimum orbital distance of detectable fragments. For the 
adopted resolution of our simulated ALMA images 
of $0.1^{\prime\prime}$, the fragments can be detected at distances down to 
50~AU. At smaller distances, 
the fragments usually merge with the central density peak. 
The likelihood for detecting the fragments reduces significantly for a lower resolution of 
$0.5^{\prime\prime}$.
Some of the most massive fragments, regardless of their orbital 
distance, can produce characteristic peaks at $\approx 5\mu$m and hence 
their presence can be indirectly inferred from the observed 
spectral energy distributions of protostars.
\end{abstract}

\section{Introduction}
Observations of young protostellar discs and numerical modeling of disc formation and evolution 
both suggest that potostellar discs may be sufficiently massive to undergo gravitational 
fragmentation 
in the course of their evolution. Observations of young protostars embedded in
parental cloud cores yield disc masses that sometimes exceed $0.1~M_\odot$ \citep[e.g.][]{Isella09,Jorgensen09}.
Numerical hydrodynamics simulations indicate that such massive discs are gravitationally
unstable and may fragment if disc cooling is sufficiently fast 
\citep[e.g.][]{Johnson03, Stamatellos09, VB2010a, Meru2011, Boss2012,Vorobyov2013}.  

The newly-formed fragments are in the sub-stellar mass regime and may be considered
as embryos of giant planets and brown dwarfs, which may or may not evolve into finished objects.
The majority of the fragments are torqued into the disc inner regions and probably onto the
burgeoning protostar, triggering episodic accretion and luminosity bursts 
\citep{VB06,VB2010a}, a phenomenon supported by many 
indirect lines of evidence \citep[see e.g.][]{DV2012}.
Some of the fragments may be ejected via many-body gravitational interaction with other fragments
into the intracluster medium, forming upon contraction a population
of freely-floating brown dwarfs \citep{BV2012}, or settle onto stable orbits providing a gateway
for the formation wide-orbit gas giant and brown dwarf companions to low-mass stars 
\citep{VB2010b,Boss2011,Boss2012,Vorobyov2013}.
Finally, fragments may loose their atmospheres via tidal striping and reveal
solid terrestrial or icy giant cores if the dust sedimentation timescale is sufficiently fast
\citep[][]{Boley10,Nayakshin10}.

Although the feasibility of disc fragmentation has been confirmed by many 
theoretical and numerical studies, unambiguous observational detection of the fragments
has been a challenging task. Recent observations using aperture masking interferometry 
have apparently been
successful in identifying one such candidate \citep{Kraus2012}, although it might have already
collapsed to a planetary-sized object as implied by the estimated age of the central star of 2~Myr.
On the other hand, numerical hydrodynamic simulations seem to indicate that fragments with typical sizes of several tens of AU can be detected with the IRAM Plateau de Bure interferometer  \citep{Stamatellos2011}.
In addition, the spiral structure in gravitationally unstable discs and fragmentation on scales
of the order of thousands AU in starless and protostellar cores
is likely to be detectable with the Atacama Large Millimeter/sub-millimeter Array (ALMA) \citep{Cossins10,Offner12}.

%With typical sizes of several tens of AU 
%\citep{Vorobyov2013}, the fragments may be directly resolved by ALMA. However, this task 
%still remains technically 
%difficult due to associated costs and uncertainty in the selection of objects.

In this paper, we study numerically the formation and evolution of a 
protostellar disc subject to vigorous gravitational fragmentation in the embedded phase
of star formation. We focus on the properties of the fragments and explore possibilities of 
their direct and indirect observational detection. In Section~\ref{model}, an overview of the
numerical model is given. Section~\ref{seds} describes the adopted model for calculating the spectral
energy distributions (SEDs). The main properties of the fragments are presented in Sections~\ref{gravinst}
and \ref{Fragprop}. The model SEDs are discussed in Section~\ref{modelSEDs} and the synthetic ALMA images
are presented in Section~\ref{alma}. The limitations of our model are discussed 
in Section~\ref{caveats} and main conclusions are given in Section~\ref{conclude}.

\section{Model description and basic equations}
\label{model}
\subsection{Numerical hydrodynamics model}
Our numerical hydrodynamics model for the formation and evolution of a star plus disc system 
is described in detail in 
\citet{VB2010a} and is briefly reviewed below for the reader's convenience.
We start our numerical simulations from the gravitational collapse of a {\it starless} cloud core, 
continue into the embedded phase of star formation, during which
a star, disc, and envelope are formed, and terminate our simulations in the T Tauri phase,
when most of the envelope has accreted onto the forming star/disc system.
%In the EPSF, the disc occupies the innermost region of our numerical grid, while the 
%larger outer part of the grid is taken up by the infalling envelope, 
%the latter being the remnant of the parent cloud core. 
In our numerical hydrodynamics simulation, the protostellar disc is not isolated 
but is exposed to intense mass loading from the envelope.  
In addition, the mass accretion rate onto the disc is not a free parameter of the model 
but is self-consistently determined by the gas dynamics in the envelope, which in turn is set
by initial conditions in the core.

To avoid too small time steps, we introduce a ``sink cell'' at $r_{\rm sc}=7$~AU and 
impose a free inflow inner boundary condition
and free outflow outer boundary condition so that that the matter is allowed to flow out of 
the computational domain but is prevented from flowing in. 
The sink cell is dynamically inactive; it contributes only to the total gravitational 
potential and secures a smooth behaviour of the gravity force down to the stellar surface.
We monitor the gas surface density in the sink cell and 
when its value exceeds a critical value for the transition from 
isothermal to adiabatic evolution, we introduce a central point-mass object.
In the subsequent evolution, 90\% of the gas that crosses the inner boundary 
is assumed to land onto the central object plus the sink cell. 
The other 10\% of the accreted gas is assumed to be carried away with protostellar jets. 
We assumed a 10\% efficiency as the typical lower limit in the X-wind model of 
jet launching \citep{Shu94}.

%\subsection{Basic equations}
We make use of the thin-disc approximation to compute the gravitational collapse of rotating, 
gravitationally unstable cloud cores. This approximation is an excellent means to calculate
the evolution for many orbital periods and many model parameters and its justification is discussed
in \citet{VB2010a}.
The basic equations of mass, momentum, and energy transport  are
\begin{equation}
\label{cont}
\frac{{\partial \Sigma }}{{\partial t}} =  - \nabla_p  \cdot 
\left( \Sigma \bl{v}_p \right),  
\end{equation}
\begin{eqnarray}
\label{mom}
\frac{\partial}{\partial t} \left( \Sigma \bl{v}_p \right) &+& \left[ \nabla \cdot \left( \Sigma \bl{v_p}
\otimes \bl{v}_p \right) \right]_p =   - \nabla_p {\cal P}  + \Sigma \, \bl{g}_p + \\ \nonumber
& + & (\nabla \cdot \mathbf{\Pi})_p, 
\label{energ}
\end{eqnarray}
\begin{equation}
\frac{\partial e}{\partial t} +\nabla_p \cdot \left( e \bl{v}_p \right) = -{\cal P} 
(\nabla_p \cdot \bl{v}_{p}) -\Lambda +\Gamma + 
\left(\nabla \bl{v}\right)_{pp^\prime}:\Pi_{pp^\prime}, 
\end{equation}
where subscripts $p$ and $p^\prime$ refers to the planar components $(r,\phi)$ 
in polar coordinates, $\Sigma$ is the mass surface density, $e$ is the internal energy per 
surface area, 
${\cal P}$ is the vertically integrated gas pressure calculated via the ideal equation of state 
as ${\cal P}=(\gamma-1) e$ with $\gamma=7/5$,
$Z$ is the radially and azimuthally varying vertical scale height
determined in each computational cell using an assumption of local hydrostatic equilibrium,
$\bl{v}_{p}=v_r \hat{\bl r}+ v_\phi \hat{\bl \phi}$ is the velocity in the
disc plane, $\bl{g}_{p}=g_r \hat{\bl r} +g_\phi \hat{\bl \phi}$ is the gravitational acceleration 
in the disc plane, and $\nabla_p=\hat{\bl r} \partial / \partial r + \hat{\bl \phi} r^{-1} 
\partial / \partial \phi $ is the gradient along the planar coordinates of the disc. 
%The planar components of the divergence of the stress tensor 
%$(\nabla \cdot \mathbf{\Pi})_p$, symmetrized velocity gradient tensor $(\nabla \bl{v})_p$, viscous %heating
%$\left(\nabla \bl{v}\right)_{pp^\prime}:\Pi_{pp^\prime}$, and 
%symmetric dyadic $\Sigma \bl{v}_p \otimes \bl{v}_p$ are found
%according to the usual rules \citep{VB10b}.

Turbulent viscosity is taken into account via the viscous stress tensor 
$\mathbf{\Pi}$, the expression for which is provided in \citet{VB2010a}.
We parameterize the magnitude of kinematic viscosity $\nu$ using a modified form 
of the $\alpha$-prescription 
\begin{equation}
\nu=\alpha \, c_{\rm s} \, Z \, {\cal F}_{\alpha}(r), 
\end{equation}
where $c_{\rm s}^2=\gamma {\cal P}/\Sigma$ is the square of effective sound speed
calculated at each time step from the model's known ${\cal P}$ and $\Sigma$. The 
function ${\cal F}_{\alpha}(r)=
2 \pi^{-1} \tan^{-1}\left[(r_{\rm d}/r)^{10}\right]$ is a modification to the usual 
$\alpha$-prescription that guarantees that the turbulent viscosity operates 
only in the disc and quickly reduces to zero beyond the disc radius $r_{\rm d}$.
In this paper, we use a spatially and temporally 
uniform $\alpha$, with its value set to $5\times 10^{-3}$. 
This choice is based on the work of \citet{VB09}, who studied numerically the long-term
evolution of viscous and self-gravitating disks. They found
that the temporally and spatially
averaged $\alpha$ in protostellar discs should lie in the $10^{-3}$-–$10^{-2}$ limits. 
Smaller values
of $\alpha$ ($\le 10^{-4}$) have little effect on the resultant mass accretion
history (dominated in this case by gravitational torques), while larger values 
($\alpha > 10^{-1}$) destroy circumstellar disks during less than 1.0~Myr of evolution and 
are thus inconsistent with mean disk lifetimes of the order of 2-–3 Myr.

%based on our recent work \citep{VB09b}. 
%Our adopted value of $\alpha$ are in agreement with the mean value inferred by \citet{Andrews09} for
%a large sample of protostellar discs {\bf in the Ophiuchus star-forming region.}

%Equation~(\ref{energ}) for the internal energy (per surface area) transport includes 
%compressional heating ${\cal P}\left( \nabla_p \cdot \bl{v}_p \right)$, radiative cooling 
%$\Lambda$, heating due to stellar/background irradiation $\Gamma$, 
%and viscous heating $(\nabla \bl{v})_{pp^\prime}:\Pi_{pp^\prime}$. 
The radiative cooling $\Lambda$ in equation~(\ref{energ}) is determined using the diffusion
approximation of the vertical radiation transport in a one-zone model of the vertical disc 
structure \citep{Johnson03}
\begin{equation}
\Lambda={\cal F}_{\rm c}\sigma\, T_{\rm mp}^4 \frac{\tau}{1+\tau^2},
\end{equation}
where $\sigma$ is the Stefan-Boltzmann constant, $T_{\rm mp}={\cal P} \mu / R \Sigma$ is 
the midplane temperature of gas\footnote{This definition of the midplane temperature is accurate within
a factor of unity \citep{Zhu2012}}, $\mu=2.33$ is the mean molecular weight, $R$ is the universal 
gas constant, and ${\cal F}_{\rm c}=2+20\tan^{-1}(\tau)/(3\pi)$ is a function that 
secures a correct transition between the optically thick and optically thin regimes. 
%cooling function (from both surfaces of the disc) 
%in the optically thick regime $\Lambda_{\rm thick}=16\, \sigma \, T^4/3\tau$ 
%and the optically thin one $\Lambda_{\rm thin}=2\,\sigma \,T^4\,\tau$.  
We use frequency-integrated opacities of \citet{Bell94}.
The heating function is expressed as
\begin{equation}
\Gamma={\cal F}_{\rm c}\sigma\, T_{\rm irr}^4 \frac{\tau}{1+\tau^2},
\end{equation}
where $T_{\rm irr}$ is the irradiation temperature at the disc surface 
determined by the stellar and background black-body irradiation as
\begin{equation}
T_{\rm irr}^4=T_{\rm bg}^4+\frac{F_{\rm irr}(r)}{\sigma},
\label{fluxCS}
\end{equation}
where $T_{\rm bg}$ is the uniform background temperature (in our model set to the 
initial temperature of the natal cloud core)
and $F_{\rm irr}(r)$ is the radiation flux (energy per unit time per unit surface area) 
absorbed by the disc surface at radial distance 
$r$ from the central star. The latter quantity is calculated as 
\begin{equation}
F_{\rm irr}(r)= \frac{L_\ast}{4\pi r^2} \cos{\gamma_{\rm irr}},
\label{fluxF}
\end{equation}
where $\gamma_{\rm irr}$ is the incidence angle of 
radiation arriving at the disc surface at radial distance $r$ calculated 
as \citep{VB2010a}
\begin{equation}
\cos{\gamma_{\rm irr}}=\cos{\alpha} \, \cos{\delta} \left(\tan{\alpha} -\tan{\delta}\right),
\end{equation} 
where
\begin{eqnarray}
\cos{\alpha}&=&{dr \over \sqrt{dr^2 +dZ^2} }; \,\, \cos{\delta}={r \over \sqrt{r^2 + Z^2}}, \\
\tan{\alpha}&=&{dZ \over dr}; \,\, \tan{\delta}={Z \over r}.
\end{eqnarray}

The stellar luminosity $L_\ast$ is the sum of the accretion luminosity $L_{\rm \ast,accr}=G M_\ast \dot{M}/2
R_\ast$ arising from the gravitational energy of accreted gas and
the photospheric luminosity $L_{\rm \ast,ph}$ due to gravitational compression and deuterium burning
in the star interior. The stellar mass $M_\ast$ and accretion rate onto the star $\dot{M}$
are determined self-consistently during numerical simulations via the amount of gas passing through
the sink cell. The stellar radius $R_\ast$ is calculated using an approximation formula of \citet{Palla91},
modified to take into account the formation of the first molecular core.
The photospheric luminosity $L_{\rm \ast,ph}$ is taken from the pre-main 
sequence tracks for the low-mass stars and brown dwarfs calculated by \citet{DAntona97}. 
More details are given in \citet{VB2010a}.

Equations~(\ref{cont})--(\ref{energ}) are solved using the method of finite differences with a time
explicit solution procedure in polar coordinates $(r, \phi)$ on a numerical grid with
$512 \times 512$ grid zones. The advection is treated using a third-order-accurate 
piecewise parabolic interpolation scheme of \citet{CW84}.
The update of the internal energy per surface area 
$e$ due to cooling $\Lambda$ and heating $\Gamma$
is done implicitly using the Newton-Raphson method of root finding, complemented by the bisection method
where the Newton-Raphson iterations  fail to converge. 
The viscous heating and force terms in equations~(\ref{mom}) and (\ref{energ}) are 
implemented in the code  using an explicit finite-difference
scheme, which is found to be adequate for $\alpha\la 0.01$.
The radial points are logarithmically spaced.
The innermost grid point is located at the position of the sink cell $r_{\rm sc}=7$~AU, and the 
size of the first adjacent cell varies in the 0.07--0.1~AU range depending on the cloud core 
size.  This corresponds to the radial resolution of $\triangle r$=1.1--1.6~AU at 100~AU.

\subsection{Initial setup}
For the initial distribution of the gas surface density 
$\Sigma$ and angular velocity $\Omega$ in the pre-stellar cores we take those 
from \citet{Basu97}
\begin{equation}
\Sigma={r_0 \Sigma_0 \over \sqrt{r^2+r_0^2}}\:,
\label{dens}
\end{equation}
\begin{equation}
\Omega=2\Omega_0 \left( {r_0\over r}\right)^2 \left[\sqrt{1+\left({r\over r_0}\right)^2
} -1\right].
\label{omega}
\end{equation}
These radial profiles are typical of pre-stellar cores formed as a result of the slow 
expulsion of magnetic field due to ambipolar diffusion, with the angular
momentum remaining constant during axially-symmetric core compression.
Here, $\Omega_0$ and $\Sigma_0$ are the angular velocity and gas surface
density at the disc centre and $r_0 =\sqrt{A} c_{\rm s}^2/\pi G \Sigma_0 $
is the radius of the central plateau, where $c_{\rm s}$ is the initial sound speed in the core. 
The gas surface density distribution described by equation~(\ref{dens}) can
be obtained (to within a factor of unity) by integrating the 
three-dimensional gas density distribution characteristic of 
Bonnor-Ebert spheres with a positive density-perturbation amplitude A \citep{Dapp09}.
In all models the value of $A$ is set to 1.2. We note that the properties
of protostellar discs formed as a result of the gravitational collapse of pre-stellar cores
depend weekly on the form of the initial density and angular velocity distributions
in the core as long as the initial core mass and ratio of rotational to gravitational 
energy $\beta$ are the same \citep{Vorobyov12}. The initial temperature of pre-stellar
cores is set to 10~K.

\subsection{Constructing spectral energy distributions}
\label{seds}
To construct  a spectral energy distribution (SED) of our model object, we split it into 
three constituent parts: the accreting protostar, 
the inner disc, and  the outer dynamic 
disc\footnote{The outer dynamic disc may include an infalling envelope, if 
present.  Because it is often difficult to draw a clear distinction between 
the disc and envelope, we refer to both simply as the outer disc.}.
This partitioning is motivated by the specifics of our numerical model---we do not 
compute the dynamical evolution of the system in the inner several AU (inside the sink cell)
due to severe Courant limitations on the hydrodynamical time step.

The basic assumption entering our calculation of the SEDs is that
our model object is viewed through an outflow cavity. This allows us to neglect the reprocessing of
light by the envelope and use surface densities and effective temperatures derived directly from
numerical hydrodynamics modelling of the outer disc and analytic model of the inner disc.
At the same time, this assumption imposes limitations on the viewing angle, 
which should not exceed the opening  semi-angle of the outflow cavity $\theta_{\rm c}$.
Below, we explain in more detail the main ideas and equations
involved in calculating the model SEDs.

\subsubsection{The outer dynamic disc}
Numerical hydrodynamics simulations provide us with gas surface densities $\Sigma$ and 
midplane temperatures $T_{\rm mp}$ in each grid cell $(i,j)$ of the outer dynamic disc.
We use a solution to the one-dimensional 
radiation transfer equation 
to calculate the radiative flux density $F_{\nu}$ leaving each grid cell of the disc surface
%each grid zone with surface area $S_{i,j}$
\begin{equation}
F^{\rm disc}_{\nu}  = {S \over d^2} B_\nu(T_{\rm surf}) 
\left(1- e^{-\Sigma \, \kappa_\nu \sec\gamma_{\rm incl}} 
\right), 
\label{CellFlux}
\end{equation}
where $B_\nu(T_{\rm surf})$ is the Plank function, $d$ is the distance to to the considered
object, set to 100~pc in the paper, 
%$\omega$ is the normal 
%angle between a specific grid cell of the  disc surface and the line of sight, 
$\gamma_{\rm incl}$ is the inclination angle of the disc
with respect to the observer (i.e., an angle between the disc rotation axis and the line 
of sight), and $S$ is the projected area of the disc surface seen by the observer. 
The inclination angle $\gamma_{\rm incl}$ is set to zero in the present paper. The effect 
of non-zero inclination,
but not exceeding the opening semi-angle of the outflow cavity, is discussed in Section~\ref{caveats} and
will be considered in detail in a follow-up paper.
%which does not exceed the 
%opening angle of the outflow cavity $\theta_{rm c}$, 
%and the expression for $\omega$ in terms of $\gamma$ and the position angle $\phi$ of 
%an individual grid cell can be found in the Appendix.
The total flux is found by summing inputs from all individual grid cells of the disc,
which amount to $512\times512$ zones for the typical numerical resolution of our hydro
simulations.

We adopt frequency-dependent opacities $\kappa_\nu$
from \citet{OH1994}. These opacities were derived for three cases of a) thick ice mantles,
b) thin ice mantles, and c) no ice mantles. In this work, we adopt opacities derived for the
case of thin ices, commonly referred to as ``OH5'' opacities in the literature. The dependence of our results on the adopted opacity
is discussed in more detail in Section~\ref{caveats}. 
%Following \citet{Bell94} we assume that
%ice mantles evaporate at $T_{\rm mp}$=210~K and use the opacities for thin/no ices 
%below/above this temperature. 

To calculate the Plunk function $B_\nu$ in equation~(\ref{CellFlux}), we need to know the disc 
surface temperature $T_{\rm surf}$.  Stellar/background irradiation of the disc surface, 
viscous/shock heating of the disc interiors, and blackbody cooling from the disc surface determine
the temperature balance of the disc in our model. Therefore, we use the following 
equation to calculate the surface temperature of the disc in our model
\begin{equation}
T^4_{\rm surf} = {1\over 2}{\cal F}_{\rm c} T^4_{\rm mp} {1 \over 1+\tau} + T^4_{\rm bg} +
{F_{\rm irr}(r) \over \sigma},
\label{Tsurf}
\end{equation}
where the factor $1/2$ is introduced because only the fluxes reaching/leaving 
one side of the disc are considered. We also note that the effect of disc opacity is now taken
into account via the factor $1/(1+\tau)$ rather than via $\tau/(1+\tau^2)$. The latter
expression yields too low temperatures in the optically thin limit $\tau \ll 1.0$
in cases when the heating sources are absent.

\subsubsection{The central accreting protostar}
The central protostar provides a significant contribution to the SED at shorter wavelengths.
The radiative flux density from the protostar $F_\nu^\ast$ is calculated assuming a black-body 
radiation spectrum with effective temperature $T_{\rm eff}^\ast$ 
% defined
\begin{equation}
F_\nu^\ast= {\pi R_\ast^2 \over d^2} B_\nu(T_{\rm eff}^\ast),
\end{equation}
where $R_\ast$ is the radius of the protostar and $T_{\rm eff}^\ast$ is defined by both 
the accretion and photospheric luminosities ($L_{\rm \ast,accr}$ and $L_{\rm \ast,ph}$, 
respectively) as
\begin{equation}
T_{\rm eff}^\ast =\left( {L_{\rm accr} + L_{\rm ph}  \over 4 \pi R_\ast^2 \sigma} \right)^{1/4},
\end{equation}
where $\sigma$ is the Stefan-Boltzmann constant. The accretion luminosity is
calculated from the model's known accretion rate $\dot{M}$ onto the protostar,
while the photospheric luminosity is taken from the pre-main 
sequence tracks of \citet{DAntona97} for low-mass stars and brown dwarfs.

\subsubsection{The inner disc}
The properties of the disc in the inner several AU cannot be calculated self-consistently
because of the sink cell, which is introduced for numerical convenience to avoid too small
timesteps near the origin where the grid lines converge. 
Nevertheless, the inner disc may contribute significantly to the total SED.
We reconstruct the structure of the inner disc using the azimuthally-averaged gas surface density profile
in the outer disc and adopting a simple model for the surface temperature of the inner disc.

The reconstruction of the physical properties of the inner disc proceeds in three steps.
First, we calculate the inner disc truncation radius $R_{\rm in}$ where 
the gas temperature exceeds the dust sublimation threshold $T_{\rm d.s.}$ = 1500~K (\citet{Dullemond01}).
The effect of varying $T_{\rm d.s.}$ is discussed in Section~\ref{caveats}.
The calculation is based on the assumption of energy balance between the energy absorbed by 
spherical, blackbody dust grains $\pi a^2 (L_{\ast,accr}+L_{\rm \ast,ph}) / (4 \pi r^2)$ and energy emitted 
by dust grains $\sigma T^4 4 \pi a^2$, where $a$ is the radius of dust grains \citep{Dunham10}. 
The resulting
expression for the sublimation radius is 
\begin{equation}
\label{Subrad}
R_{\rm in}=\sqrt{\frac{L_{\rm \ast,accr} + L_{\rm \ast,ph}}{16 \pi \sigma T_{\rm d.s.}^4}}.
\end{equation}
We assume that the inner rim of the disc at $R_{\rm in}$ is heated to the temperature of
dust sublimation $T_{\rm d.s.}$.

Second, we calculate the gas surface density distribution in the inner disc 
using the following relation 
\begin{equation}
\label{SigmaSink}
  \Sigma_{\rm in}(r)=\Sigma_{\bf jn,0} \left( \frac{r}{r_{\rm jn}} \right)^{-p},
\end{equation}    
where $r_{\rm jn}$ is the outer radius of the inner disc and $\Sigma_{\rm jn,0}$ 
is the corresponding gas surface density. To make a smooth transition 
between the inner and outer discs, we 
perform the least-squares fitting to the azimuthally averaged gas surface density of the latter.
Based on this fitting, we choose a value of $\Sigma_{\rm jn,0}$ at $r_{\rm jn}$ 
such that the two discs are smoothly joined.  
The exponent $p$ varies between 1.5 for gravitationally unstable discs and $\approx 1.0$ 
for viscosity-dominated discs \citep[e.g.][]{VB09}. 
Since the inner disc is likely to be gravitationally stable, we chose $p=1.0$.
%The inner and outer discs are joined together at $r_{\rm jn}=20$~AU, 
%the latter value is chosen so as 
%to make a smooth transition between the two discs.
%The gas surface density of the inner disc $\Sigma_0$ at $r_{\rm jin}$ is found using 
%the least-squares fitting of the
%azimuthally averaged gas surface density in the outer dynamic disc. 
We note however that variations in the density profile of the inner disc 
insignificantly influence the resulting SED due to the 
high optical depth of the inner disc.

Finally, the surface temperature of the inner disc is found adopting a simple model
of a viscous disc in Keplerian rotation. Viscosity generates dissipation 
of energy in the vertical column of the disc at a rate (erg~cm$^{-2}$~s$^{-1}$)
\begin{equation}
\dot{E}={9\over 4} \nu \Sigma \Omega^2, 
\end{equation}
where $\Omega=(G M_\ast/r^3)^{1/2}$.

Assuming further that the inner disc radiates from its surface as a black-body, 
one can write the disc surface temperature as 
\begin{equation}
T^4_{\rm surf}= {9 \over 8 \sigma} \nu \Sigma \Omega^2 + T^4_{\rm bg} + {F_{\rm irr} 
\over \sigma}. 
\end{equation}
The inner disc scale height $Z_{\rm in}$, needed to calculate the incident radiation 
flux $F_{\rm irr}$, is found by assuming the disc aspect ratio from \citet{DAlessio99}
\begin{equation}
{Z_{\rm in} \over r}=C\left( {r \over r_{\rm jn}} \right)^{0.25},
\end{equation}
where the scaling factor $C$ is found by applying the disc scale height of the outer dynamic 
disc at $r_{\rm jn}$. This ensures a continuous transition between the inner and outer 
disc vertical scale heights. For a steady-state disc, the dynamic viscosity 
$\nu \Sigma$ can be expressed in 
terms of the mass accretion rate $\dot{M}$ as \citep{Pringle81}
\begin{equation}
\nu \Sigma = {\dot{M} \over 3 \pi} \left( 1- \left({R_\ast \over r}\right)^{1/2}  \right).
\end{equation}
For the mass accretion rate $\dot{M}$ in the inner disc we adopted the value 
calculated at the inner inflow boundary of the outer dynamic disc, $r_{\rm sc}=7$~AU.
Once the surface temperature and density in the inner disc have been calculated, 
the input to the total SED is found using equation~(\ref{CellFlux}).
The implementation of a more accurate, time-dependent model 
of the inner disc, subject to mass loading from the outer dynamic disc, 
remains to be a future improvement of our numerical model.

\section{Disc instability and fragmentation}
\label{gravinst}

Theoretical and numerical studies of the evolution of protostellar
discs indicate that disc fragmentation is a complicated
phenomenon, which can be influenced by both the internal disc
physics and external environment. The latter may influence
the disc susceptibility to fragmentation directly through, e.g.,
stellar and background irradiation, \citep[e.g.][]{VB2010a,Stamatellos2011} or 
indirectly by setting the initial conditions in
cloud cores that favor or disfavor fragmentation in subsequently
formed discs due to, e.g., higher/lower infall rates onto the disc \citep{VB06,Kratter08,Vor2011}. 

\begin{figure*}
 \centering
  \includegraphics[width=13cm]{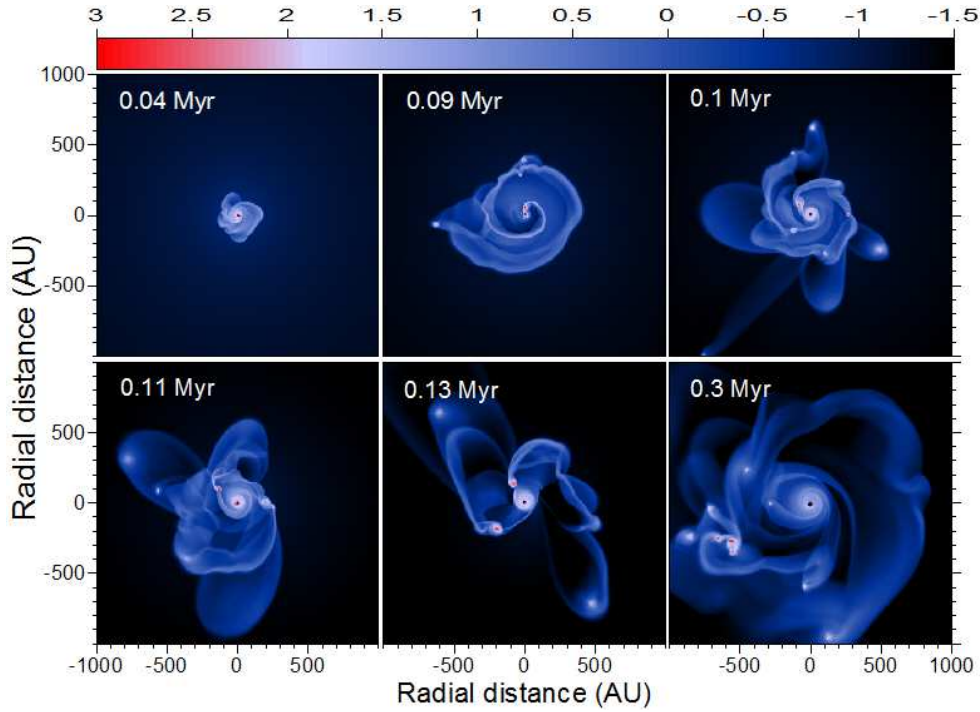}
%  \resizebox{\hsize}{!}{\includegraphics{figure1.eps}}
  \caption{Gas surface density maps of the inner $2000\times2000$~AU box in our model. The time elapsed since
  the formation of the central protostar is indicated in each panel. The scale bar is in log
  g~cm$^{-2}$. The protostar is at the coordinate centre.}
  \label{fig1}
\end{figure*}

In this section, we describe the time evolution of protostellar discs
formed as a result of the gravitational collapse of  pre-stellar cores with different initial mass 
and angular momentum. We consider two prototype cases: model~1 with initial core mass
$M_{\rm c}=1.22~M_\odot$ and ratio of rotational to gravitational energy $\beta=8.8\times10^{-3}$,
and also model~2 with $M_{\rm c}=0.74~M_\odot$ and $\beta=2.75\times10^{-3}$. 
We have chosen the two models based on our previous numerical simulations \citep{VB2010a} 
so as to obtain discs with vigourous and occasional gravitational 
fragmentation.

Figure~\ref{fig1} presents the gas surface density images for model~1
in the inner $2000\times2000$~AU box at 
various times since the formation of the central protostar. 
The entire computational box is almost 10 times larger. 
The disc forms at $t=0.01$~Myr and becomes gravitationally unstable at $t=0.04$~Myr. 
A first set of fragments appear in the disc at $t=0.1$~Myr. 
The subsequent evolution is characterized by vigourous gravitational instability and
fragmentation leading sometimes to a nearly complete breakup of the disc into 
massive fragments and dense filamentary arms (i.e., at $t=0.13$~Myr). 
Some of the fragments possess mini-discs of their own, complicating even more the dynamics of the
entire disc-like structure. 

\begin{figure*}
 \centering
  \includegraphics[width=13cm]{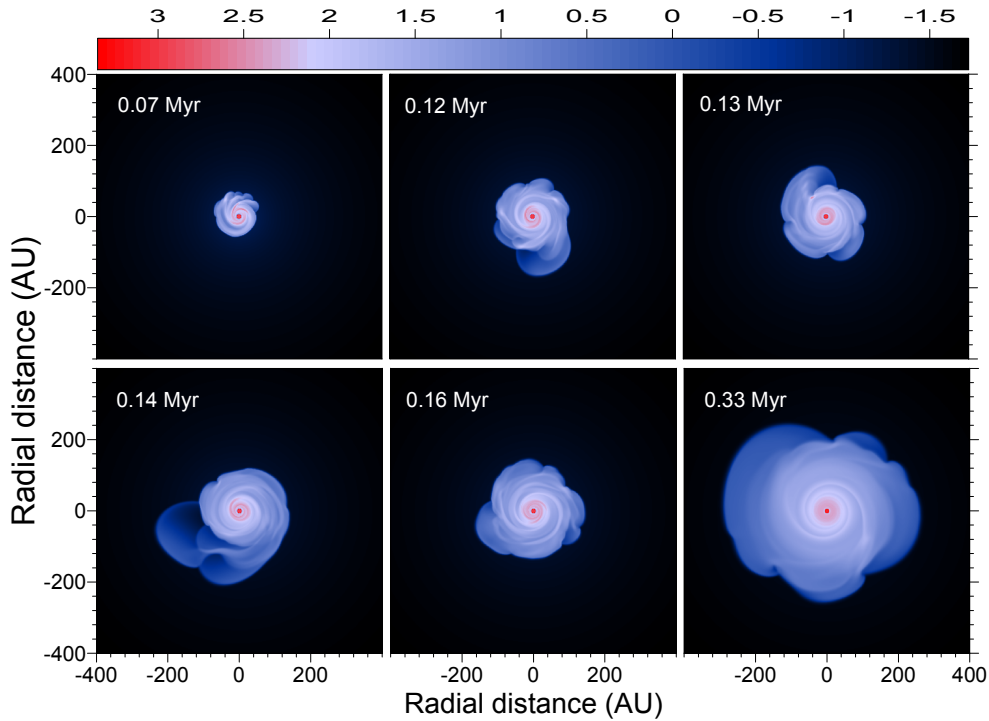}
%  \resizebox{\hsize}{!}{\includegraphics{figure6.eps}}
  \caption{The same as Fig.~\ref{fig1} only for model~2.}
  \label{fig2}
\end{figure*}

When studying the susceptibility of a protostellar disc to gravitational fragmentation, the 
following two criteria are usually investigated \citep[e.g.][]{Rafikov05,VB2010a}. \\
(i) The ratio of the local cooling time $t_c = e/\Lambda$ 
to the local dynamical time $\Omega^{-1}$ should be smaller than a few
\citep{Gammie01,Rice03,Meru2012}, with the actual value 
depending on the physical conditions in the disc.
Following \citet{VB2010a}, we will refer to the dimensionless quantity
$t_{\rm c} \Omega^{-1}$ as the ${\cal G}$-parameter and adopt 
${\cal G} = 1$ as a fiducial (and conservative) critical value. \\
(ii) The Toomre criterion $Q =c_{\rm s} \Omega/(\pi G \Sigma)$ 
for a Keplerian disc is smaller than some critical value $Q_{\rm cr}$, 
usually taken to be unity \citep{Toomre64}. The value of $Q_{\rm cr}$ may depend
on the physical conditions and may vary by a factor of
unity \citep{PPS97}. We have chosen a conservative value of $Q_{\rm cr}=1.0$.  

We applied these two criteria to our model at $t=0.13$~Myr in a fashion similar to
figures 3 and 4 in \citet{VB2010a}. We found that the fragments obey both criteria.
At the same time, there are a few regions in the disc that are characterized by $Q<1.0$ but with 
no obvious fragmentation taking place in there. The lack of fragmentation in these regions is 
explained by slow cooling.
We also note that both criteria for disc fragmentation are violated in the inner several tens 
of AU, in agreement with previous studies of fragmenting discs.

We note that the total mass of the disc (including fragments and 
the sink cell) varies between $0.19~M_\odot$ at $t=0.09$~Myr and
$0.25~M_\odot$ at $t=0.6$~Myr, with the maximum value of $0.31~M_\odot$ reached at $t=0.25$~Myr.
The mass of the central star grows from $0.32~M_\odot$ at $t=0.09$~Myr to $0.76~M_\odot$ at $t=0.6$~Myr.
For the disc mass estimates, we used an algorithm described in \citet{DV2012} and a
critical surface density for the disc to envelope transition of 0.1~g~cm$^{-2}$. 

Finally, in Figure~\ref{fig2} we present an example of a gravitationally unstable disc 
showing only occasional fragmentation. The panels show the gas surface density maps in model~2
for the inner $800\times800$~AU box. The time elapsed since the formation of the central protostar
is indicated in each panel. 
The disc in this model forms at $t=0.035$~Myr, appreciably later than in
model~1 due to smaller angular momentum of the parental pre-stellar core. 
The resulting disc mass is smaller by almost a factor of two than in model~1.
The disc is gravitationally unstable and exhibits
a flocculent spiral structure but is mainly stable to fragmentation, 
in contrast to model~1. The only distinct episode
of disc fragmentation takes place around $t=0.13$~Myr, but the fragment has 
quickly migrated through the inner sink cell during just several orbital periods. 
In the subsequent evolution, the disc shows little signs of fragmentation.
We note that the disc age 
in each panel is approximately equal to that in the corresponding panel in Figure~\ref{fig1}.

\section{Properties of the fragments}
\label{Fragprop}
\begin{figure}
  \resizebox{\hsize}{!}{\includegraphics{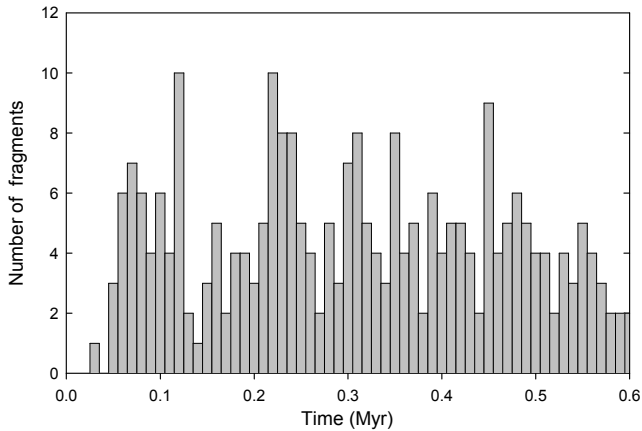}}
  \caption{Number of fragments vs. time in model~1. The number of fragments at a given time instant
  is calculated using the fragment tracking algorithm described in \S~\ref{Fragprop}.
  An increase in the number of fragments shows recent fragmentation,
  and a decrease shows recent destruction/accretion of the fragments.}
  \label{fig3}
\end{figure}

In this section, we explore the properties of the fragments in model~1 with 
the purpose of determining the feasibility of using 
observations to indirectly or directly detect their presence.
Figure~\ref{fig3} shows the number of fragments $N_{\rm f}$ calculated 
every 0.01~Myr after the formation of the protostar 
using a fragment tracking algorithm described
later in this section. Evidently, $N_{\rm f}$ varies significantly with time, reaching local peak values
every $\approx0.1$~Myr. 
An increase in the number of fragments shows recent fragmentation,
and a decrease shows recent destruction/accretion of the fragments.

It is interesting to note that the time intervals between recurrent peaks 
is of the same duration as the characteristic time of mass infall 
onto the disc
\begin{equation}
t_{\rm infall} = {M_{\rm d} \over \dot{M}_{\rm d}},
\end{equation}
where $M_{\rm d}$ is the disc mass and $\dot{M}_{\rm d}\propto c_{\rm s}^3 /G $ 
is the mass infall rate onto the disc. Indeed, for the disc mass 
of 0.2--0.25~$M_\odot$ and mass infall rates of $(2-3)\times10^{-6}~M_\odot$, typical for our model,
the characteristic infall time ranges between 0.07~Myr and 0.12~Myr. The approximate 
match between periodicity of the peaks in Figure~\ref{fig3} and $t_{\rm infall}$ 
suggests that the mass infall may cause periodic bursts of disc fragmentation, at least in 
the embedded phase of disc evolution which lasts for 0.4-0.45~Myr in this model.
The important implication of this phenomenon is that the fragments can be present
in the disc for a significantly longer time period than expected for just one burst of disc fragmentation
in numerical hydrodynamics simulations without disc infall 
\citep[e.g.][]{Stamatellos2011b}, significantly increasing the probability for observing the fragments.

We use a clump tracking mechanism described in \citet{Vorobyov2013} to 
determine the position and properties of the fragments in model~1. The method is briefly outlined below.
Fragments are identified on the computational mesh based on two criteria 
\begin{eqnarray}
\label{pres}
{\partial {\cal P} \over \partial r^\prime} &+& {1 \over r^\prime}{\partial {\cal P} \over \partial \phi^\prime} <0, \\
{\partial \Phi \over \partial r^\prime} &+& {1 \over r^\prime}{\partial \Phi \over \partial \phi^\prime} >0,
\label{grav}
\end{eqnarray}
where $\Phi$ is the gravitational potential and 
the primed coordinates are defined as $r^\prime=r-r_{\rm c}$ and $\phi^\prime=\phi-\phi_{\rm c}$.
The polar coordinates $r_{\rm c}$ and $\phi_{\rm c}$ denote the geometrical centre of the fragment,
 which is
determined as the position of a local maximum in the gas surface density on the computational mesh.
The first condition 
mandates that the fragment must be pressure supported, with a negative pressure gradient
with respect to the centre of the fragment. The second condition requires that the fragment
is kept together by gravity, with the potential well being deepest at the centre of the fragment.
All grid cells that satisfy both equations~(\ref{pres}) and (\ref{grav}), starting from the first layer
of cells immediately adjacent to ($r_{\rm c},\phi_{\rm c}$), are 
defined as belonging to the fragment. 

\begin{figure}
  \resizebox{\hsize}{!}{\includegraphics{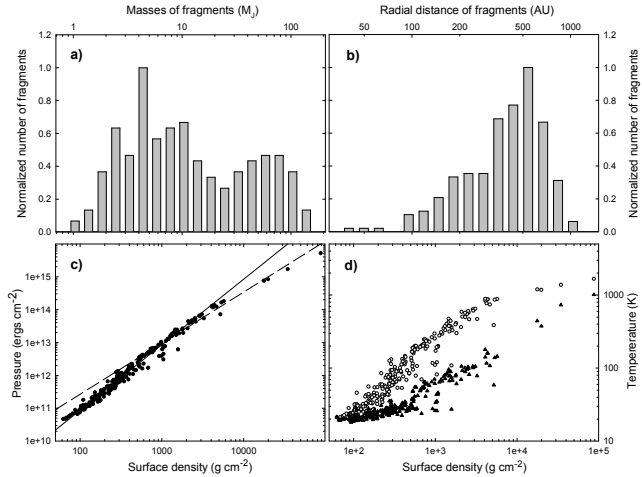}}
  \caption{Properties of fragments in model~1. {\bf Panel a)}---normalized 
  distribution function of masses of the fragments. {\bf Panel b)}---normalized distribution 
  function of radial positions of the fragments. {\bf Panel c)}---vertically 
  integrated gas pressure ${\cal P}_{\rm c}$ vs. gas surface density $\Sigma_{\rm c}$ 
  at the geometrical centre of the fragments. 
  {\bf Panel d)}---Midplane temperatures $T_{\rm mp,c}$ (open circles) and 
  surface  temperatures $T_{\rm surf,c}$ (filled triangles)
  vs. $\Sigma_{\rm c}$ at the centers of the fragments. The lines are
  the best least-squares fits to the data with $\Sigma_{\rm c}<10^3$~g~cm$^{-2}$ (solid line)
  and $\Sigma_{\rm c}\ge 10^3$~g~cm$^{-2}$ (dashed line). The corresponding exponents are 2.0 and 1.5.}
  \label{fig4}
\end{figure}

Panels a) and b) in Figure~\ref{fig4} present the normalized distribution 
functions of fragment mass and radial positions in model~1.
These distributions were calculated by searching for and identifying
the fragments in the inner 5000~AU of the computational mesh 
every 0.01~Myr starting from $t=0.03$~Myr and ending at $t=0.6$~Myr. 
%After 1.0~Myr of disc 
%evolution, only one fragment survived and seemed to settle onto a quasi-stable orbit. 
The obtained masses and radial positions of the fragments were then split into 20 
logarithmically spaced bins. One can identify two local maxima in the mass distribution function, 
one belonging to the 
planetary-mass regime at $\approx 5~M_{\rm Jup}$ and the other lying in the 
upper brown-dwarf-mass regime at $\approx 60-70~M_{\rm Jup}$. The whole mass spectrum extends 
from about a Jupiter mass
to very-low-mass stars. The radial positions of the fragments have a well-defined maximum at 
$\approx 400-500$~AU. The number of the fragments gradually declines at smaller distances and 
no fragments are seen at $r<30$~AU. The dearth of fragments at $r<100$~AU is most likely 
explained by slow cooling that impedes fragmentation in the inner several tens 
AU \citep{Johnson03,Meru2011} 
and fast inward radial migration of the fragments.  Numerical simulations show that once migration 
has started it proceeds very fast and fragments spiral into the inner few AU 
(and probably onto the star) 
during just a few orbital periods \citep{VB06,VB2010a,Machida2011,Cha11}, making it difficult
to observe such fragments.  We note that we do not follow
the fate of the fragments once they have passed through the sink cell at 7~AU.

The two bottom panels in Figure~\ref{fig4} present vertically integrated pressures ${\cal P}_{\rm c}$,
surface densities $\Sigma_{\rm c}$, midplane temperatures $T_{\rm mp,c}$ and surface 
temperatures $T_{\rm surf,c}$ derived  at the geometrical centers of the fragments
in model~1.
%We intentionally did not perform any sort of averaging over 
%the volume occupied by a given fragment in order to analyze the maximum values of aforementioned quantities.
As before, these data were calculated by searching for and identifying fragments 
in the inner 5000~AU every 0.01~Myr starting from $t=0.03$~Myr and ending at $t=0.6$~Myr.
We emphasize that the derived data do not always correspond to distinct fragments but
sometimes represent the same fragment at different evolutionary times.
The solid and dashed lines in panel~c) of Figure~\ref{fig4} present 
the least squares fitting to the ${\cal P}_{\rm c}$ vs $\Sigma_{\rm c}$ data.
In particular, the solid line provides the best fit for the fragments with 
$\Sigma_{\rm c}<10^3$~g~cm$^{-2}$, while the dashed line does that for the fragments with 
$\Sigma_{\rm c}\ge10^3$~g~cm$^{-2}$. The resulting exponents are 
2.0 (solid line) and $1.55$ (dashed lines), implying that the fragments can be described by
a polytrope with the polytropic index $n$=(1.0--2.0) (related to the adiabatic index as $\gamma=(n+1)/n$).
The value of $n$ seems to depend on the density.
Radiative fragments are supposed to have n=$1.5$, while convective ones are characterized by $n=2.5$.
Since in our 2D approach the vertical convection is not taken into account, it is not surprising
that our fragments are radiative. However,  mixing in the disc plane may 
still take place for the most massive fragments, which may explain 
values of the polytropic index greater than 1.5.

The open circles in panel d) of Figure~\ref{fig4} show midplane temperature vs. surface 
density, while the
filled triangles show surface temperatures vs. surface density. The surface temperatures
are appreciably lower than the midplane ones, as can be expected for optically thick fragments.
We have not identified any 
fragments with $T_{\rm mp,c}$ exceeding 2000~K, implying that all the fragments are the first
hydrostatic cores which have not yet dissociated molecular hydrogen. 

\begin{figure}
  \resizebox{\hsize}{!}{\includegraphics{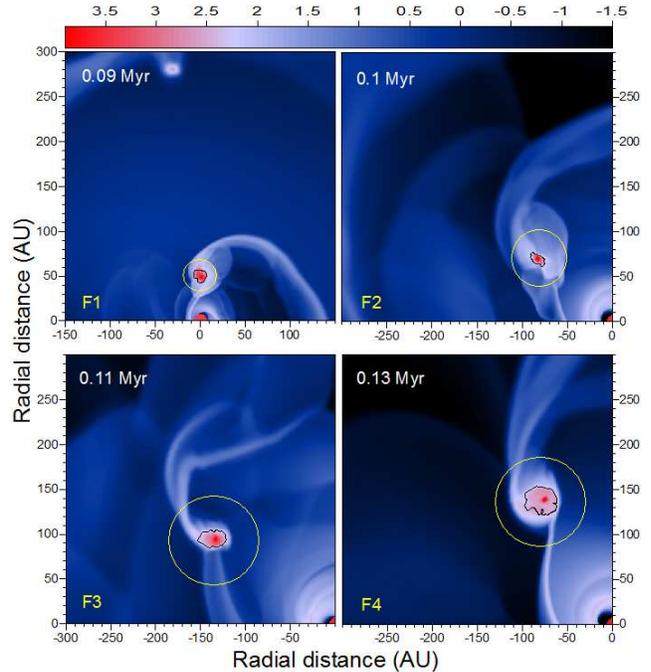}}
  \caption{Zoomed-in surface density images of four fragments in model~1 characterized by 
  midplane temperature exceeding $10^3$~K. Black curves outline the fragments as 
  found by our fragment tracking algorithm. Yellow circles mark the corresponding Hill radii. 
  Time elapsed since the formation of the central protostar
  is indicated in each panel. The scale bar is in g~cm$^{-2}$.}
  \label{fig5}
\end{figure}

Typical disc surface densities and temperatures at $r\ga100$~AU, where most fragments are located, 
do not, as a rule, exceed 100~g~cm$^{-3}$ and 50~K, respectively \citep{Vor2011}. 
Most fragments in our model are characterized by surface densities greater than 100~g~cm$^{-2}$. 
%which may facilitate their observational detection. 
Surface temperatures of the fragments
are usually much lower than those at the midplane due to the 
high optical depths of the fragments.
The majority of the fragments are characterized by  $T_{\rm surf,c}<100$~K.
%There is only a handful of fragments with $T_{\rm surf,c}$ exceeding $100$~K. 
However, there are four fragments
with midplane temperatures exceeding $10^{3}$~K, which means that they have started evaporating dust
in their interiors. This leads to a significant drop in opacity and an associated increase in 
the surface temperatures reaching several hundred Kelvin or even higher.

In Figure~\ref{fig5} we present surface density images, 
zoomed in on the four fragments in model~1 with midplane temperatures $T_{\rm mp,c}$ 
exceeding $10^3$~K. 
The black curves outline the fragments as identified by our tracking
mechanism. The yellow circles mark the Hill radii for each fragment defined as
\begin{equation}
R_{\rm H}= r_{\rm f} \left( {1\over 3} {M_{\rm f} \over M_\ast + M_{\rm f}} \right)^{1/3},
\end{equation}
where $r_{\rm f}$ is the radial distance from the protostar
to the fragment and $M_{\rm f}$ is the mass of the fragment confined
within the black curve. 
%The fragments are surrounded by 

The properties of the four fragments are listed in Table~\ref{table2}. 
In particular, the columns from left to right present: 
1) the number of the 
fragment, 2) the time at which the fragment is seen in the disc, 3) the mass 
of the fragment,4) the mass confined within the Hill radius $M_{\rm H}$, 5) the 
radial distance to the fragment, 6) the radius of the fragment $R_{\rm f}$, 
7) the Hill radius of the fragment, 8) the midplane temperature of the 
fragment, 9) the surface temperature of the fragment, 10) the surface density 
of the fragment, and 11) the optical depth to the midplane $\tau$. The last 
four quantities were calculated at the geometrical centers of the fragments. 
The radius of the fragment is found by calculating the surface area occupied 
by the fragment and assuming that the fragment has a circular form. 

Evidently, all four fragments with $T_{\rm mp,c}\ge 10^3$~K are in the brown-dwarf-mass  regime.
In fact, we do not find any planetary-mass objects with midplane temperatures exceeding 400~K.
Fragments that are closer to the protostar are generally smaller in size, probably due to tidal stripping.
We do not see any significant correlation between $T_{\rm mp,c}$ and $r_{\rm f}$ for the entire sample
of the fragments in  model~1, implying that hot fragments may be found at any distances and 
not only at $r<170$~AU as in the current model.
The last column in Table~\ref{table2} illustrates a strong dependence of opacity on
midplane temperature. For fragment~F2 with $T_{\rm mp,c}$=1180~K the opacity is $\tau=256$,
while for fragment~F1 with $T_{\rm mp,c}=1660$~K the opacity drops to $\tau=18.5$.
This drop is caused by evaporation of dust grains and results in a significant increase
in the surface temperature, exceeding $10^3$~K for the hottest fragment.
A more detailed study of the properties of the fragments 
will be presented in a follow-up paper.
%No fragments with $T_{\rm c}\ge 10^3$~K were seen beyond 200~AU,  

\begin{table*}
\begin{center}
\caption{Characteristics of the fragments}
\label{table2}
\renewcommand{\arraystretch}{1.5}
\begin{tabular}{c c c c c c c c c c c }
\hline \hline
fragment &time & $M_{\rm f}$ & $M_{\rm H}$ & $r_{\rm f}$ & $R_{\rm f}$ & $R_{\rm H}$ 
& $T_{\rm mp}$ & 
$T_{\rm surf}$ & $\Sigma$  & $\tau$ \\
& (Myr) &  ($M_{\rm Jup}$) & ($M_{\rm Jup}$) & (AU) &  (AU) & (AU)  & (K) & (K) &  (g~cm$^{-2}$) &   \\ 
\hline %[-2.0ex]
F1 & 0.09 & 55 & 68 & 48  & 8  & 17 & 1660 & 1020 & $8.7\times10^4$ & 18.5  \\
F2 & 0.1  & 32 & 47 & 108 & 9  & 32 & 1180 & 380  & $2.0\times10^4$ & 256  \\
F3 & 0.11 & 52 & 63 & 163 & 16 & 56 & 1190 & 445  & $1.8\times10^4$ & 139  \\
F4 & 0.13 & 64 & 84 & 158 & 21 & 55 & 1375 & 740  & $3.5\times10^4$ & 32  \\
%5 & 1.55 & 1.27 & 0.75 & 27.5 & 40.5 & 180 & 0.06 & 19.5 & 41 & 3.9 \\
%6 & 1.4 & 0.56 & 0.95 & 3.5 & 5.0 & 190 & 0.05 & 11.5 & 20.5 & 4.1 \\
\hline
\end{tabular}
\end{center}
\medskip
$M_{\rm f}$--the mass of the fragment, $M_{\rm H}$--the mass confined within the Hill radius, 
$r_{\rm f}$--the radial distance to the fragment, $R_{\rm f}$--the radius of the fragment, 
$R_{\rm H}$--the Hill radius of the fragment, $T_{\rm mp}$--the midplane temperature of the 
fragment, $T_{\rm surf}$--the surface temperature of the fragment, 
$\Sigma$--the surface density of the fragment, and 
$\tau$--the optical depth to the midplane of the fragment.

\end{table*}

\section{Model spectral energy distributions}
\label{modelSEDs}

\begin{figure}
  \resizebox{\hsize}{!}{\includegraphics{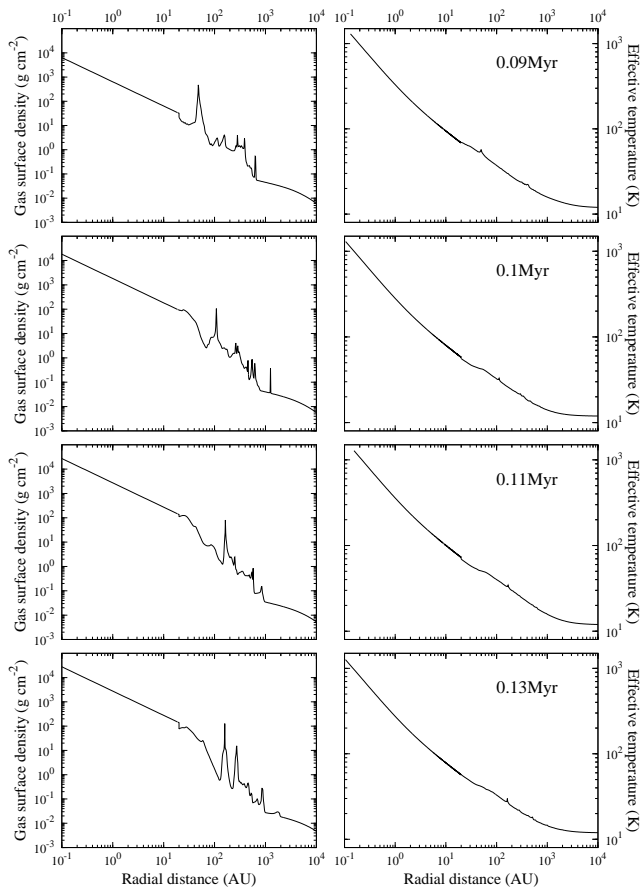}}
  \caption{Azimuthally averaged gas surface density (left-hand column) and effective temperature
  (right-hand column) distributions in our model at four times after the formation of the central protostar.}
  \label{fig6}
\end{figure}

In this section, we calculate model SEDs using the method laid out
in Section~\ref{seds}. Our object is split into three constituent parts:
the protostar, the inner disc, and the outer dynamic disc (plus envelope if present).
%To determine the input from the central protostar, we use 
%accretion luminosity known from our modelling and photospheric luminosity taken from 
%the pre-main sequence tracks of \citet{DAntona97}. The input from the outer 
%disc is calculated based on the model's known gas surface densities and effective temperatures
%found from equation~(\ref{Tsurf}). The contribution from the inner disc (sink cell) is
%computed based on the gas surface density distribution extrapolated
%down to the dust sublimation radius. The effective temperature of the inner disc
%is found by applying a simple model of an irradiated viscous disc.
%First, we focus on evolutionary times when hot and massive fragments are present in model~1.
%These fragments are shown in Figure~\ref{fig5} and their properties are listed
%in Table~\ref{table2}. 
Figure~\ref{fig6} shows the azimuthally averaged gas surface 
density ($\overline\Sigma$; left panels) and surface temperature 
($\overline{T}_{\rm surf}$; right panels) profiles in model~1 at the four times when the 
most massive fragments are present in the disc. 
The inner and outer discs are joined together at $r_{\rm j}=20$~AU. 
%The time (indicated in each panel) was chosen so as to focus on hot and massive fragments.
By assumption, the radial profiles of $\overline\Sigma$ and $\overline{T}_{\rm surf}$ 
in the inner disc are axisymmetric, but they are not in the outer disc---the averaging 
procedure could not smooth out azimuthal inhomogeneities caused by spiral arms and fragments.
The most massive fragments manifest themselves by clear-cut peaks in $\overline\Sigma$ 
at $r\approx 50$~AU 
($t=0.09$~Myr), $r\approx110$~AU ($t=0.1$~Myr), and $r\approx 160$~AU ($t=0.11$~Myr and $t=0.13$~Myr).
There are also a number of secondary peaks corresponding to less massive fragments. 
$\overline{T}_{\rm surf}$ exceeds $10^3$~K near the dust sublimation radius 
($R_{\rm in}\approx 0.1-0.15$~AU, depending on the total luminosity) and declines gradually to
10~K in the envelope. There are also local maxima seen in $\overline{T}_{\rm surf}$ at the positions
of the fragments but their magnitudes are considerably smaller than those of 
%The azumthally averaged $T_{\rm eff}$ also exhibits local maxima at the positions of the fragments
%but their magnitudes are considerably lower 
$\overline\Sigma$ due to 
substantial input of isotropic stellar irradiation into the thermal balance of the outer disc.

\begin{figure}
  \resizebox{\hsize}{!}{\includegraphics{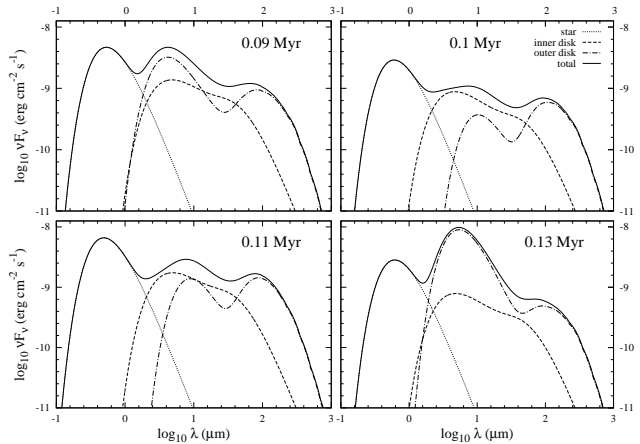}}
  \caption{Model spectral energy distributions in model~1 at four times after the formation of the central
  star.  Individual contributions from 
the central protostar, inner disc, and outer disc  are shown by the dotted, 
dashed, and dash-dotted lines, respectively. The solid line is the
total flux.}
  \label{fig7}
\end{figure}

Figure~\ref{fig7} presents the resulting SEDs in model~1. In particular, individual 
contributions from 
the central protostar, inner disc, and outer disc  are shown by the dotted, 
dashed, and dash-dotted lines, respectively. The solid line is the
total flux $\nu F_{\rm \nu}$ in units of erg~cm$^{-2}$~s$^{-1}$. The distance to our model
object is taken to be 250~pc. 
%The opening semi-angles of the outflow cavities calculated using
%equation~(\ref{opening}) are $51^\circ$ ($t=0.09$~Myr), $53^\circ$ ($t=0.1$~Myr),
%$56^\circ$ ($t=0.11$~Myr), and $61^\circ$ (t=0.13~Myr).  
%Large opening angles of the outflow cavity increase the chances of
%detecting the fragments.

A visual inspection of Figure~\ref{fig7} reveals that hot and massive fragments
may contribute significantly to the total radiative flux. In particular, 
the contribution from the outer disc has a characteristic double-peaked structure, 
with the peak at shorter wavelengths produced by the most massive and hot fragments in the disc. 
The input from the fragments to the total flux  may sometimes exceed that from 
the inner disc (dotted lines).  As a result, the total flux may  
have a characteristic secondary peak at $\lambda\approx 5~\mu$m comparable 
in magnitude to that produced by the protostar at $\lambda\approx 0.5~\mu$m.
This feature is clearly visible at $t=0.09$~Myr and $t=0.13$~Myr when fragments F1 and F4 
have midplane temperatures exceeding 1300~K. At these high 
temperatures dust grains start to evaporate, which leads to a sharp drop in opacity and
the corresponding increase in the surface temperature. As Table~\ref{table2}
demonstrates, $T_{\rm surf}$  exceeds 700~K for fragment F1 at $t=0.09$~Myr and fragment F4 at 
$t=0.13$~Myr. In comparison, fragment F2 at $t=0.1$~Myr and fragment F3 at $t=0.11$~Myr are 
characterized by midplane temperatures that are
below the dust sublimation temperature of $\approx 1200$~K. Their surface temperatures
are  below 450~K, their input to the total SEDs is comparable to that of the inner 
disc, and the secondary peak at $\lambda\approx 5~\mu$m is less evident.
%The opacities are still rather
%high and the corresponding surface temperatures are comparable to those of the inner disc.

\begin{figure}
  \resizebox{\hsize}{!}{\includegraphics{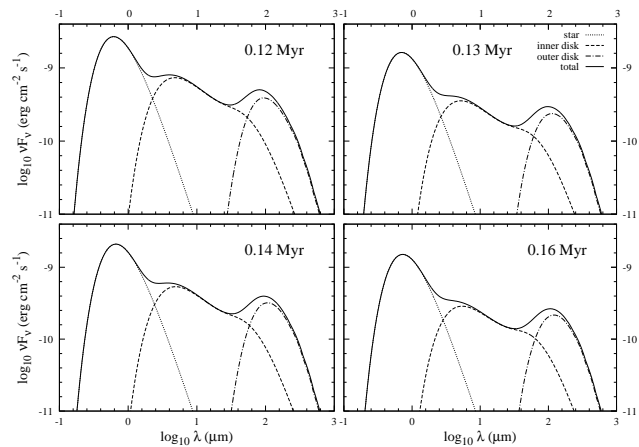}}
  \caption{Same as for Figure~\ref{fig7} only for model~2.}
  \label{fig8}
\end{figure}

To highlight the predicted SED differences between fragmenting and non-fragmenting
discs, we show in Figure~\ref{fig8} the spectral energy distributions 
for the non-fragmenting model~2. Four ages of the disc that approximately match those 
of model~1 in Figure~\ref{fig7} were considered.  Evidently, the SED of the outer disk  
(dash-dotted lines)
has a single-peaked form with a maximum at $\approx 100\mu$m. The outer disk contributes 
a negligible fraction to the total flux at $\lambda\approx 5~\mu$m, unlike the fragmenting model~1
showing a secondary peak at this wavelength. {\bf  Nevertheless, there is a small feature at 
$\lambda\approx 5~\mu$m, which is not related to the outer disc and is likely caused 
by simplistic modelling of the inner disc
due to the limitations of our numerical hydrodynamics scheme. }

%lack of self-consistent hydrodynamic treatment of the inner disk regions

Several recent studies have presented observations and models of 
the SEDs of relatively large samples of embedded protostars and T Tauri stars.  
For example, \citet{Robitaille07,Furlan08,Furlan11} all presented such data 
and models for targets in Taurus.  No triple-peaked SEDs like those seen at 
0.09, 0.11, or 0.13 Myr in Figure~\ref{fig7} are seen in these studies, 
suggesting that such massive fragments are not present most of the time.  
Indeed, such massive fragments should only last at most a few thousand years 
before migrating onto the star, approximately 1\% of the embedded stage 
duration of a few $\times 10^5$ yr \citep{Evans09,Vor10,DV2012}.  Even if several 
fragments form and migrate inward during the embedded stage, the total time 
when such features are observable is likely less than 10\%, especially 
considering the necessity of favorable viewing geometry.  Thus it is not 
surprising that no triple-peaked SEDs like those shown in Figure~\ref{fig7} 
have been detected given that studies like those listed above generally only 
contain a few dozen embedded sources.  Should such a feature be detected in 
the future, it would be a strong indication of a massive disk fragment.

\section{Synthetic ALMA images of fragmenting discs}
\label{alma}
In the previous section, we have shown that the most massive and hot fragments 
can leave characteristic signatures in the SEDs of young protostellar discs
observed through outflow cavities.
In this section, we explore the possibility of detecting the fragments directly
using interferometric observations with the Atacama Large Millimeter/sub-millimeter Array (ALMA).

To investigate this possibility, we first generated 
synthetic images following the same procedures described above for producing 
SEDs except, instead of summing over all spatial positions, we calculated the 
intensity in each pixel in a cartesian coordinate grid with a grid spacing 
of 5 AU (0.025$''$ at the assumed distance of 250 pc).  We then used the 
{\it simobserve} and {\it simanalyze} tasks in the CASA software 
package\footnote{CASA, the Combined Astronomy Software Applications, is the 
ALMA data reduction and analysis package; see http://casa.nrao.edu/ for 
details.} to simulate 1 hour of ALMA integration at both 230 and 345 GHz, 
where these particular frequencies were chosen as the best compromise between 
resolution, sensitivity, and availability of suitable weather for such 
observations.  The antenna configuration most closely corresponding to a 
synthesized beam of 0.1$''$ (25 AU at the assumed distance of 250 pc) was 
selected for each frequency, since such a resolution is the minimum necessary 
to adequately resolve the disc into its individual fragments.  With these 
simulated observing parameters, the 3$\sigma$ detection limits of the ALMA 
images are approximately 0.2 mJy beam$^{-1}$ at 230 GHz and 0.8 mJy beam$^{-1}$ 
at 345 GHz. 

\begin{figure*}
 \centering
  \includegraphics[width=17cm]{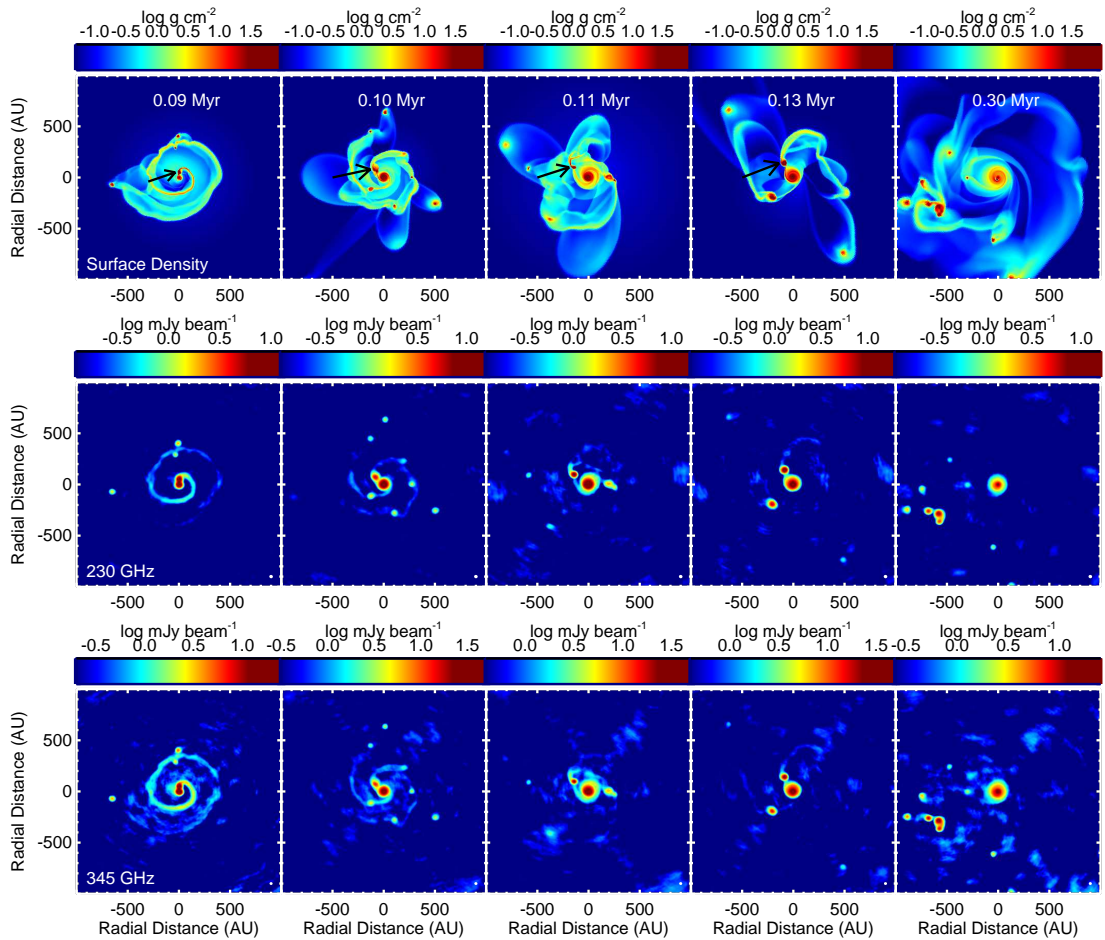}
%  \resizebox{\hsize}{!}{\includegraphics{figure9.eps}}
  \caption{{\it Top row:} Gas surface density maps of the 
inner $2000\times2000$~AU at five times since the formation of the central 
protostar. The black arrows point to the most massive fragments F1--F4.  
{\it Middle and bottom rows:} The corresponding synthetic ALMA images at 
230~GHz (middle) and 345~GHz (bottom). The beam size is indicated in white in 
the bottom right of each panel.  The images are displayed on a logarithmic 
scaling to capture the full dynamic range of the central density peaks, 
surrounding fragments, and spiral disc structure.}
  \label{fig9}
\end{figure*}  

\begin{figure*}
 \centering  
  \includegraphics[width=17cm]{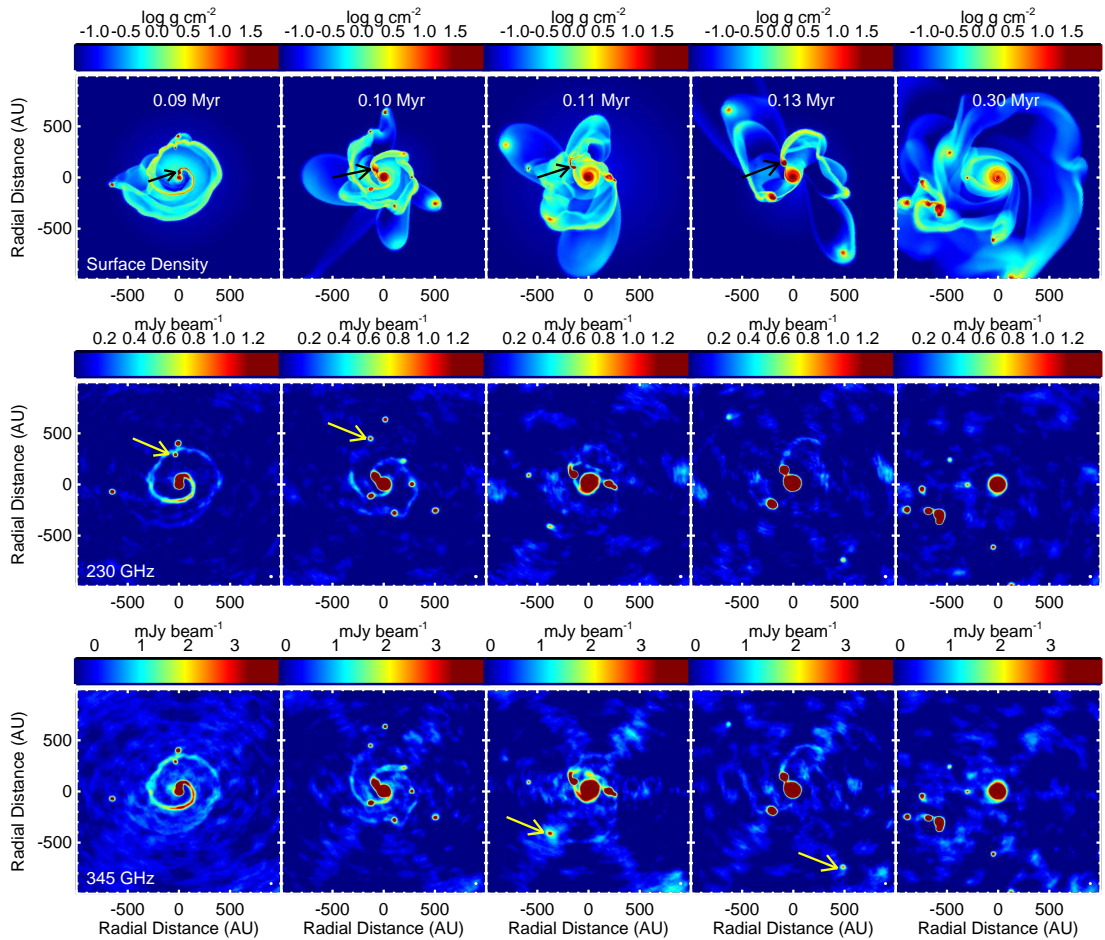}  
%    \resizebox{\hsize}{!}{\includegraphics{figure10.eps}}
  \caption{Same as Figure~\ref{fig9}, except the images are 
displayed on a linear scaling with the scale ranges chosen to emphasize the 
detections of the disc fragments above the disc spiral structure, at 
the expense of oversaturation of the central density peaks.  The yellow arrows 
highlight the least massive fragments clearly detected in the images.}
  \label{fig10}
\end{figure*}

Figure~\ref{fig9} presents the gas surface densities in model~1 (top row, log g~cm$^{-2}$)
and simulated ALMA images at 230~GHz (middle row) and 345~GHz (bottom row), both in log mJy/beam.  
The actual input images to the ALMA data 
simulator are flux density images as described above, not surface density 
images, but these input images are qualitatively very similar to the surface 
density images shown in the top row. Both 230 and 345 GHz are 
sufficiently in the Rayleigh-Jeans limit such that the emission is only 
linearly dependent on temperature and thus primarily traces surface density, 
which varies over a much larger range than the temperature. 
The  time elapsed since the formation of the protostar 
is indicated in the top row and  the beam size 
is plotted in the right-bottom corner of each image. The black arrows in the top row
point to the most massive fragments shown in Figure~\ref{fig5} and described in detail in Table~\ref{table2}.
%The log scaling for the ALMA images was
%chosen to highlight the fragments and especially the elements of spiral arms, since they are 
%appreciably weaker than the maximum values near the central density peaks. 

Evidently, some of the fragments can be confidently detected with ALMA using just one hour of integration
time. For instance, fragments F2--F4 at a radial distance of about 100--150~AU 
are clearly seen in the image and their peak fluxes are well
above the detection limits of 0.2 (0.8) mJy beam$^{-1}$ at 230 
(345) GHz.  
On the other hand, fragments that are located at radial distances
$r\la50$~AU start to merge with the central flux peak and are more difficult 
to detect with our chosen resolution of $0.1^{\prime\prime}$. For instance, 
fragment~F1 (at $r_{\rm f}=48$~AU) is beginning to merge with
the central peak. This is not surprising considering that the 
corresponding linear resolution is 25~AU for the adopted distance of 250~pc. 
%and angular resolution of $0.5^{\prime\prime}$, the linear resolution is about 50~AU. 
However, we note that, once ALMA beings full-science array 
operations, angular resolutions better than 0.1$''$ 
will be achievable, enabling even very closely spaced fragments to be 
separated (although longer integration times than the 1 hour considered here 
will be required to reach comparable sensitivities).  We also note that the 
spiral structure is clearly evident in some images. 
%{\color{red}and that the images at 
%345~GHz reveal fragments and the spiral structure better than those at 230~GHz
%(I think we should delete the portion of the sentence I have 
%colored in red}).

While the log scaling is useful to emphasize the spiral structure, it 
unfortunately also enhances the noise. Images with a linear
scaling chosen to oversaturate the central density peaks highlight the weak, 
distant fragments, thus Figure~\ref{fig10} presents such images for the same 
five times as in Figure~\ref{fig9}.  The maximum image intensities exceed 
20~mJy beam$^{-1}$ but we have set the maximum of the image scales to values 
in the approximate range of 1--4 mJy beam$^{-1}$, thus oversaturating the 
images by factors of three or more.  Evidently, this helps to 
resolve weaker and more distant fragments at the cost of losing fragments at 
$r_{\rm f}\la 100$~AU, which are now almost completely merged with the central 
density peaks.

The yellow arrows in the middle and bottom rows of Figure~\ref{fig10} indicate 
the least massive fragments that can be detected with ALMA.  These fragments 
have masses of $3.5~M_{\rm Jup}$, $2.5~M_{\rm Jup}$, $2.0~M_{\rm Jup}$, 
$3.5~M_{\rm Jup}$, and $1.5~M_{\rm Jup}$ at $t=0.09$~Myr, t=0.10~Myr, $t=0.11$~Myr, 
$t=0.13$~Myr, and $t=0.3$~Myr
respectively.  Obviously, ALMA can detect fragments down to the planetary mass 
regime. The largest orbital distance of detectable fragments is about 850~AU.

\begin{figure*}
 \centering  
  \includegraphics[width=17cm]{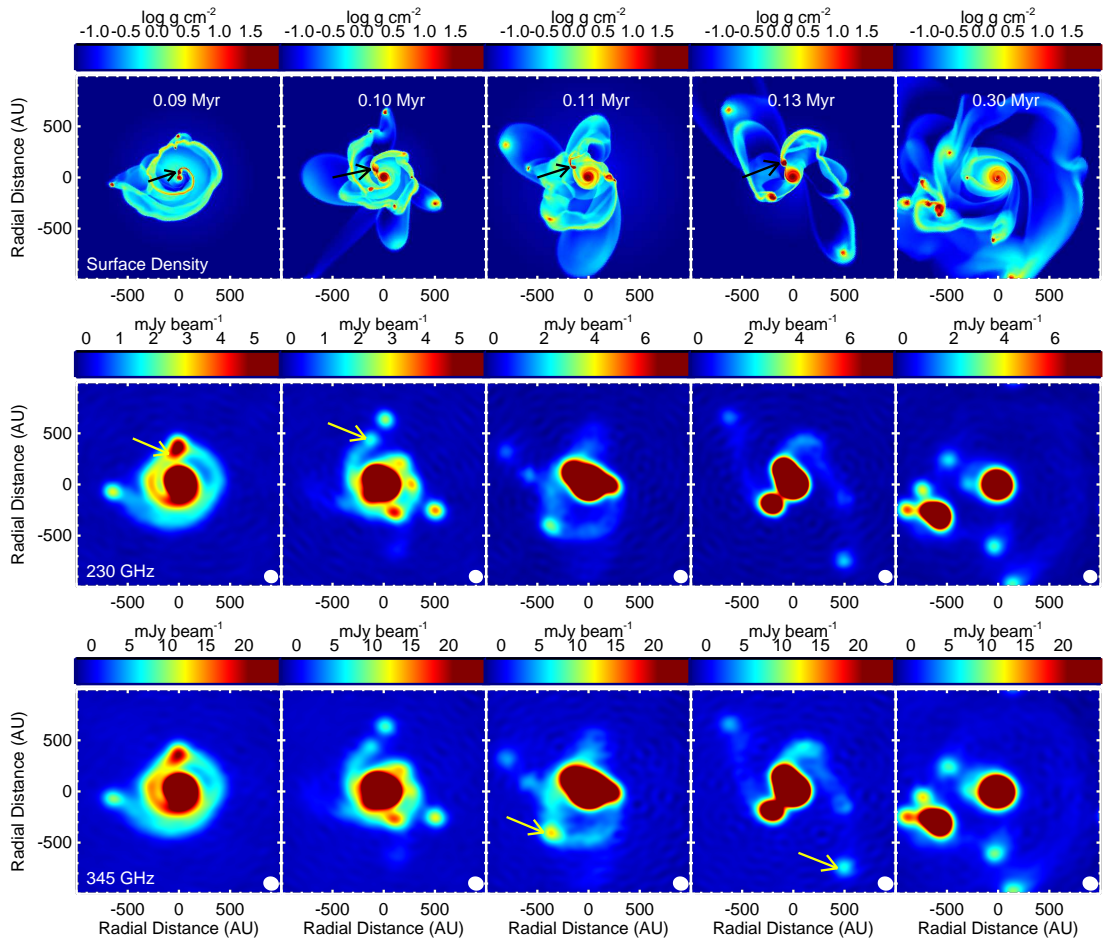}  
%    \resizebox{\hsize}{!}{\includegraphics{figure10.eps}}
  \caption{Same as Figure~\ref{fig10}, except for a reduced resolution of 0.5$''$.}
  \label{fig11}
\end{figure*}

To illustrate the effect of reduced resolution, we generated the simulated ALMA images 
in model~1 at the same five times as in Figure~\ref{fig10} but for a resolution of 0.5$''$.
The corresponding images are shown in Figure~\ref{fig11} using the same oversaturated scaling
as in Figure~\ref{fig10}. Evidently, fragments located at radial distances from the protostar 
$\la 200$~AU have completely merged with the central peak. In addition, close-separation 
fragments now appear as one fragment (e.g., at $t=0.09$~Myr and 0.3~Myr).
Consequently, the total number of identifiable fragments has reduced considerably. We conclude that
a resolution of 0.5$''$ can be used for discs located at distances $\la 150$~pc.

%345 GHz is better than 230 GHz ...

\section{Model caveats}
\label{caveats}

{\it Uncertainties in dust opacity}. 
In this study, we used the frequency-integrated opacities of \citet{Bell94} to calculate
the disc thermal balance and also the frequency-dependent
opacities of \citet{OH1994} for the case of thin ices to calculate the flux. 
Uncertainties in the former 
may affect the calculated surface temperatures and hence the shape of the model SEDs.
In particular, if dust grains evaporate at higher temperatures than assumed in \citet{Bell94}, 
than the characteristic secondary peak at $\lambda\approx 5\mu$m 
(see Fig.~\ref{fig7}) will be less pronounced.
Numerical simulations with other opacity tables are needed to assess the effect.

On the other hand, uncertainties in the adopted frequency-dependent opacities
have little effect on the resulting SEDs at $\lambda\la 100\mu$m 
due to high optical depth at these frequencies but may have some effect at longer wavelengths.
To evaluate the magnitude of this effect, we calculated the SEDs with frequency-dependent opacities
of \citet{OH1994} for the three cases of thin ice, thick ice, and no ice. The resulting 
radiative fluxes at 230 GHz and 345 GHz differ by a factor of two at most.
Finally we note that opacity is varying vertically, an effect that 
cannot be captured in our 2D approach.
%Dependence of SEDs on opacity...  

\begin{figure}
  \resizebox{\hsize}{!}{\includegraphics{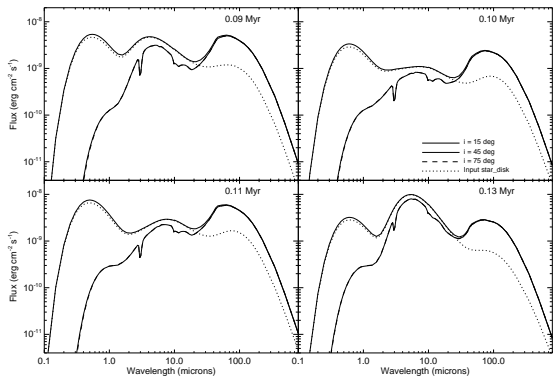}}
  \caption{Spectral energy distributions for model 1 at the same 4 
times shown in Figure~\ref{fig7} obtained from Monte Carlo radiative transfer 
models of the net star+disk system embedded in a core (see text for details).  In 
each panel, the model SEDs at three different inclinations are plotted, 
along with the star+disk SED given as an input to the models. Note that the 
$i=45^\circ$ and $i=75^\circ$ curves nearly merge due to an almost spherical core shape.}
  \label{fig12}
\end{figure}

{\it Non-zero inclination}. In this paper, we have constructed SEDs and synthetic ALMA
images using a methodology of \citet{ChG97}  and assuming zero inclination, i.e., 
for objects viewed pole-on. Our method is best applied to a later evolutionary stage 
when most of the
parental core has dissipated, thus neglecting possible photon scattering or
reprocessing by the envelope.  
To better compare with other studies that 
present synthetic SEDs of protostars based on global simulation data 
\citep[e.g., ][]{Kurosawa04,Offner12a}, we ran 
simple Monte Carlo radiative transfer models of a star+disk system embedded 
within a core. {\bf For this purpose, we used the dust radiative 
transfer package RADMC \citep{DD2004}. }
The SEDs of the net star+disk system shown in Figure~\ref{fig7} 
were adopted as the input SEDs.  A flattened, rotating
density structure following the \citet{Terebey84} solution, with an outflow 
cavity with an opening semi-angle of 30\degree\, was assumed for the core, and 
the adopted dust opacities were taken from OH5 \citep{OH1994},
 modified to include both absorption and isotropic scattering opacities (see 
\citet{Dunham10,DV2012} for details).  The results are shown in 
Figure~\ref{fig12}.  Reprocessing by the dust in the core becomes significant 
for high inclinations but is negligible for inclinations smaller than the 
semi-opening angle of the outflow cavity, and the excess emission at long 
wavelengths from dust in the core does not hinder the ability to detect the 
triple-peaked SED indicative of massive disk fragments.  Furthermore, our 
models do not suggest that scattering will significantly affect the observed 
mid-infrared SED to the point of hindering the detection of massive disk 
fragments, but we acknowledge that our models likely underestimate this 
effect since we are unable to properly treat scattering as anisotropic.

We also note that for large values of $\gamma_{\rm incl}$, the position angle 
of the fragment in the disc  becomes important. The fragment may 
be enhanced/suppressed if it is located on the farthest/nearest side of the disc 
with respect to the observer. 
%We note that according to our basic assumption the inclination angle $\gamma_{\rm incl}$ 
%should note exceed the opening semi-angle of the outflow cavity $\theta_{\rm c}$

{\it Background irradiation}. The temperature of background irradiation is set to 10~K in
the present study. A higher temperature of
background irradiation makes the overall disc warmer. The net result is that 
the Jeans length is increasing and the fragments become more massive, which may facilitate 
their detection. At the same time,
the disc becomes less prone to fragmentation and, as numerical hydrodynamics simulations
of \citet{VB06} demonstrated, fragmentation ceases at $T_{\rm bg}\ga 50$~K. 

{\it Dust sublimation temperature}. In this study, we adopted a dust sublimation temperature
of 1500~K to calculate the sublimation radius in equation~(\ref{Subrad}). 
We found that varying $T_{\rm d.s.}$ in the 1000--2000~K limits influences our results
insignificantly. In particular, for $T_{\rm d.s.}=1000$~K the maximum decrease in the flux by 18\%
(with respect to that calculated for $T_{\rm d.s.}=1500~K$) occurs at $\lambda=1.2~\mu$m, while for
$T_{\rm d.s.}=2000$~K, the corresponding increase in the flux by 46\% occurs at $\lambda=1.75~\mu$m.

{\it Numerical resolution}. 
The Truelove criterion \citep{Truelove99} states that the Jeans length has to be resolved
by at least four grid zones in order to correctly study disc fragmentation.
The Jeans length $R_{\rm J}$ of a thin, fragmenting disc can be expressed as 
\citep{Vorobyov2013}
\begin{equation}
R_{\rm J} = {\langle v^2 \rangle \over \pi G \Sigma_{\rm d}},
\label{RJeans}
\end{equation}
where $\langle v^2 \rangle = 2 {\cal R} T_{\rm d}/\mu$ is the velocity dispersion of a thin disc,
$\cal R$ is the universal gas constant, $T_{\rm d}$ is the gas midplane temperature 
in the fragment and $\Sigma_d$ is the gas surface density at the fragment-disc
interface. 
Fragments usually condense out of densest sections of spiral arms. The typical 
surface densities and temperatures in spiral arms do not exceed 100~g~cm$^{-2}$ and 
100~K \citep{Vor2011}.
Adopting these values for $\Sigma_{\rm d}$ and $T_{\rm d}$, the corresponding Jeans length 
is $R_{\rm J}\approx 20$~AU.  

In our model, the radial and azimuthal grid resolution
at $r=100$~AU is $\approx 1.0$~AU and the Jeans length is resolved by roughly 20 
grid zones in each coordinate direction, sufficient to fulfil the Truelove criterion.
The resolution deteriorates with increasing distance from the central star
and the Truelove criterion is expected to break at $r\ga 500$~AU, where
the spatial resolution on our logarithmically spaced grid starts to exceed 5.0~AU and 
the Jeans length is resolved by less than four grid zones in each coordinate direction.
However, fragmentation takes place mostly at radial distances from a few tens to a 
few hundreds AU. Fragments that are seen in our model at larger distances 
are most likely scattered from the 
inner disc regions due to gravitational interaction with other fragments\footnote{In the most
extreme cases, some of the fragments may even be ejected from the disc into 
the intracluster medium \citep{BV2012}.}. Higher resolution simulations are planned for the near future
to confirm our results.

{\it Contraction timescale}. 
The grid resolution imposes limitations on the minimum size of a fragment 
of the order of $1.0$~AU. This means that we cannot follow the contraction of fragments 
to  planetary- or sub-stellar-sized objects and hence we may overestimate the feasibility of using
(sub-)millimeter observations to detect the presence of fragments
(fully formed planetary or sub-stellar objects will become essentially invisible for the ALMA).
%This affects the feasibility of using sub-millimeter observations to detect 
%the presence of fragments in the disc, since fully formed planetary or sub-stellar objects will become %essentially invisible to the ALMA.  
The contraction timescale of a fragment to a fully formed object strongly depends on
its initial mass. A one-Jupiter-mass fragment will contract for about a few~$\times 10^5$~yr, 
while a ten times more massive fragment will collapse in $<10^4$~yr \citep{Tohline02,Helled10,Boley10}.
These timescales are comparable to or shorter than the timescale between successive bursts of disc 
fragmentation, $(7-12)\times10^{4}$~yr (see Figure~\ref{fig3}), which periodically regenerate the population
of fragments in the disc. This implies that
the probability of detecting fragments increases with decreasing mass of the fragment and it may be
quite difficult to ''catch'' a massive proto-brown dwarf in the process of formation.

{\it The effect of environment and magnetic fields.} 
{In the present study, we considered the formation of protostellar discs
from isolated non-magnetized cores. Numerical hydrodynamics simulations of star cluster formation 
indicate that radiation feedback from young stars and tidal torques can significantly
reduce the disk propensity to fragment \citep[e.g.][]{Bate09,Offner09}. Our model takes into account
disc irradiation by the central protostar and background blackbody radiation with a temperature
appropriate for isolated cores, $T_{\rm bg}=10$~K, but does not include tidal stripping of the 
disc. 
%We also set the temperature of 
%the background radiation to a value that is appropriate for isolated cores, $T_{\rm bg}=10$~K.
Our results are thus more applicable to isolated star formation in a quiescent environment. 
In addition, magnetic braking can significantly reduce
the disc mass and size in the early Class 0 phase (sometimes even preventing disc formation), 
also diminishing the disc propensity to fragment
\citep{Li11,Joos12}. The effect of magnetic fields can thus reduce the likelihood of detecting
the fragments.

\section{Conclusions}
\label{conclude}
Using numerical hydrodynamics simulations in the thin-disc limit,
we studied the formation and evolution of a gravitationally unstable protostellar disc
with a particular emphasis on the properties of fragments formed via disc gravitational
fragmentation. We constructed the spectral energy distribution of our model object viewed pole-on,
including the flux from the central protostar, the inner unresolved disc and the outer
disc plus envelope system. We also used the ALMA software and generated synthetic images of
our model disc at two frequencies best fitted for the fragment detection. We found the following.
\begin{enumerate}
%\item The protostellar disc is prone to gravitational fragmentation if both the Toomre $Q$-parameter %and 
%the $\cal G$-parameter (the product of the local cooling time and angular velocity) 
%is below unity, in agreement with many previous studies \citep[e.g.][]{Gammie01,Rice03,VB2010a}.
\item The disc experiences multiple episodes of fragmentation in the embedded phase
of star formation, with the time between recurrent fragmentation episodes approximately
equal to the characteristic time of mass infall onto the disc $M_{\rm d}/\dot{M}$.
The fragments can therefore be present
in the disc for a significantly longer time period than expected for just one burst of disc fragmentation
in numerical hydrodynamics simulations without disc infall.
%the periodicity of the bursts set by the mass infall rate onto the disc.

\item The mass distribution function of the fragments have two maxima, one in the
planetary-mass regime around $5~M_{\rm Jup}$ and the other in the upper brown-dwarf-mass
regime around $60-70~M_{\rm Jup}$. The whole mass spectrum extends from about a Jupiter mass
to very-low-mass stars. The radial positions of the fragments have a maximum at 400--500~AU,
with a dearth of fragments at $r<100$~AU caused presumably by fast inward migration at these distances.

\item The majority of the fragments are characterized by surface temperatures $T_{\rm surf}<100$~K due
to high optical depths. Occasionally, some fragments may attain temperatures in their interiors sufficient
to evaporate dust grains. Such fragments are characterized by much higher surface densities and can
leave characteristic peaks at $\approx 5\mu$m in the SEDs of young protostellar discs viewed through
the outflow cavity. These features are comparable in magnitude to the flux from the central star and
can potentially be used to infer the presence of massive ($\ga 50~M_{\rm Jup}$) and hot fragments
in protostellar discs. Non-fragmenting discs lack notable peaks around $5\mu$m. 

\item The majority of the fragments in our model can be detected with ALMA using just
one hour of integration time and a resolution of $0.1^{\prime\prime}$, provided that
the disc is favourably inclined. The detection limit
on the mass of the fragments is $1.5~M_{\rm Jup}$ at a distance of 250~pc. 
It is advisable to use a log scaling to 
resolve the spiral structure and fragments at small distances $\la 100$~AU, while the 
oversaturated linear scaling is best fitted to detect distant and low-mass fragments.
The likelihood of detecting the fragments reduces significantly for a lower resolution of
0.5$''$ but can  probably be used for objects located at distances of the order of 150~pc and
less.

In this paper, we have explored only one fragmenting model. More numerical simulations exploring 
a wide parameter space of initial core masses, the effects of different inclinations, and environments are further needed.

\end{enumerate}

\section*{Acknowledgments} 
The authors are thankful to the referee for valuable comments and suggestions that
helped to improve the manuscript.
EIV acknowledges support from the RFBR grant 11-02-92601-KO.
Numerical simulations were done 
   on the Atlantic Computational  Excellence Network (ACEnet), Shared Hierarchical
Academic Research Computing Network (SHARCNET), and Vienna Scientific Cluster (VSC-2).
OVZ acknowledges grant from the OeAD - Austrian Agency for International Cooperation in Education \& Research,
financed by the Scholarship Foundation of the Republic of Austria.

\appendix

\section{Testing the SED algorithm}
To test the spectral energy distribution algorithm described in \S~\ref{seds}, we 
tried to reproduce the model of \citet{ChG97} for a flared disc heated by stellar irradiation. 
In this model, the disc surface temperature is calculated as 
\begin{equation}
T_{\rm surf}=\left( {\alpha \over 2}  \right)^{1/4} \left(  R_\ast \over r \right)^{1/2}  T_\ast,
\end{equation}
where the grazing angle is $\alpha=0.005 r_{\rm AU}^{-1} + 0.05r_{\rm AU}^{2/7}$ and 
$r_{\rm AU}$ is the radial distance in AU.  
%This model is for passive discs that surround 
%T Tauri stars. The calculations were performed to reproduce the SEDs for a system with a flaring blackbody %disc, viewed pole-on. 
We used this temperature distribution and applied the inner and outer disc cutoff 
radii 0.07~AU and 200~AU, respectively\footnote{
We note that \citet{ChG97} used a value of 270~AU for the outer disc radius.}.
%(this is the one parameter that differs form initial data, $R_{\rm out}$ = 270AU in \citet{ChG97}),
%respectively, and equation (1) in \citet{ChG97} for the radial distribution of the effective temperature.
The star was modeled as a spherical black body with temperature $T_{\ast}$ = 4000K and 
radius $R_{\ast}$ = 2.5 $R_{\odot}$, as in \citet{ChG97}.

The comparison of the resulting SED with that presented in \citet{ChG97} is shown in Figure~\ref{fig13}.
Our SED is shown by the solid line, with the individual contributions of the disc and the star
plotted by dashed and dotted lines, respectively. The SED of \citet{ChG97} is indicated with filled
circles (stellar input) and asterisks (disc input), respectively.
Figure~\ref{fig13} indicates that our algorithm can reasonably well reproduce the results 
of \citet{ChG97}.

\begin{figure}
  \resizebox{\hsize}{!}{\includegraphics{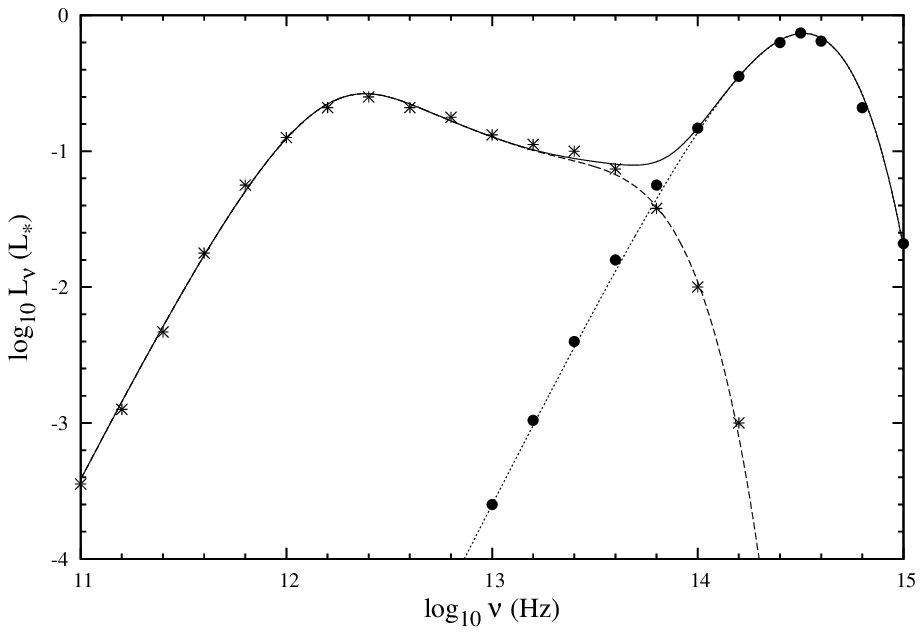}}
  \caption{Comparison of our test SED with that of \citet{ChG97}. Individual contributions 
  to the test SED from the central star and the disc are shown with dotted and dashed lines, 
  respectively; the solid line is the total flux.  The SED of Chiang \& Goldreich are plotted
  with asterisks (disc input) and filled circles (stellar input).}
  \label{fig13}
\end{figure}

\end{document}